\documentstyle[aps,epsf,epsfig,prl,psfig]{revtex}
\newcommand{\avg}[1]{\langle{#1}\rangle}
\newcommand{\be}{\begin{equation}\FL}
\newcommand{\ee}{\end{equation}}
\newcommand{\beas}{\begin{eqnarray*}}
\newcommand{\eeas}{\end{eqnarray*}}
\newcommand{\bea}{\begin{eqnarray}\FL}
\newcommand{\eea}{\end{eqnarray}}
\newcommand{\req}[1]{(\ref{#1})}
\def\sign{\hbox{sign}\,}

\begin{document}

\twocolumn[\hsize\textwidth\columnwidth\hsize\csname
@twocolumnfalse\endcsname
\title{Phase Transition and Symmetry Breaking in the Minority Game}
\author{Damien Challet$^{(1)}$ and Matteo Marsili$^{(2)}$}
\address{$^{(1)}$Institut de Physique Th\'eorique, 
Universit\'e de Fribourg, CH-1700\\
$^{(2)}$Istituto Nazionale per la Fisica della Materia,
Unit\'a di Trieste SISSA,
V. Beirut 2-4, Trieste I-34014}
\date{\today}
\maketitle
\widetext

\begin{abstract}
We show that the Minority Game, a model of interacting heterogeneous 
agents, can be described as a spin systems and it displays a phase 
transition between a symmetric phase 
and a symmetry broken phase where the games outcome is 
predicable. As a result a ``spontaneous 
magnetization'' arises in the spin formalism.
\end{abstract}

\narrowtext

]

Market interactions among economic agents give rise 
to fluctuation phenomena which are raising much
interest in statistical physics\cite{pw,ZENews}. 
The search for a toy system to study agents with 
market-like interactions has led to the definition of the Minority 
Game\cite{ZENews,CZ1} (MG), a model inspired by Arthur's ``El Farrol'' 
problem\cite{Arthur}, which embodies some basic market 
mechanisms \cite{ZENews} while keeping the mathematical complexity 
to a minimum.

In short, the MG is a repeated game where $N$ agents have to 
decide which of two actions (such as buy or sell) to make. 
With $N$ odd, this procedure identifies a {\em minority action}
as that choosen by the minority. Agents who took the minority
action are rewarded by one payoff unit, whereas the majority of 
agents looses one unit.
Agents do not communicate one with the other and they
have access to a ``public information'' -- related to past
game outcomes -- represented by one of $P$ possible patterns. 

The strategic point of view of game theory may 
require, in a case like this, a prohibitive computational 
task for each of the agents\cite{games}. 
That is specially true if $N$ and $P$ are very 
large and agents have not complete information on the detailed 
mechanism which determines their payoffs, the identity of their 
opponents or even their number $N$. In such complex 
strategic situations -- which are similar to those that agents 
face in stock markets\cite{ZENews,ZPal} -- agents may prefer to 
simplify their decision task by looking for simple behavioral 
rules which prescribe an action for each of the $P$ possible 
patterns. This may be particularly advantageous if computational 
costs exist.


This behavior, called {\em inductive reasoning} in ref. \cite{Arthur}, 
is the basis of the MG\cite{ZENews,CZ1}: 
each agent has a pool of $S$ rules 
which prescribe an action for each of the $P$ patterns. At each time, 
she follows her best rule (see below for a more precise definition).
These rules, called strategies below, are initially drawn at 
random among all possible rules, independently for each 
agent in order to model agents' heterogeneity of beliefs and
behaviors. 

Numerical simulations\cite{CZ1,Savit,johnson} 
have shown that this system displays a {\em cooperative} phase
for large values of the ratio $\alpha=P/N$:
With respect to the simple ``random agent'' state --
where each agent just tosses a coin to choose her action --
agents are better off because they get to enstablish a
sort of coordination.
For small values of $\alpha$ agents receive, on average, 
poorer payoffs than in the random agent state, a behavior
which has been related to crowd effects in 
markets\cite{ZENews,Savit,johnson}.
A qualitative understanding of this behavior has been 
given in terms of geometric considerations\cite{ZENews,CZ2}.

In this Letter we show that the model can be described
as a spin system and, as $\alpha=P/N$ varies, it undergoes 
a dynamical phase transition with symmetry breaking. 
The symmetry which gets broken is the equivalence 
between the two actions: in the symmetric phase 
($\alpha<\alpha_c$) both actions are taken by the
minority with the same frequency (e.g. there are,
on average, as many buyers as sellers). For $\alpha>
\alpha_c$, in each of the $P$ possible states, the
minority does more frequently an action than the other
one, i.e. the game's outcome is asymmetric. 
An asymmetry in the game's outcome is an opportunity that 
an agent could in principle exploit to gain. This is called 
an {\em arbitrage} in economics and it bears a particularly 
relevant meaning (see discussions in \cite{ZENews,Savit}).
The asymmetry for $\alpha>\alpha_c$ naturally suggests an order
parameter and is related to a ``phase separation'' in the population 
of agents: while for $\alpha<\alpha_c$ all agents use all of 
their strategies, for $\alpha>\alpha_c$ a finite fraction 
$\phi$ of the agents ends up using only one strategy which, 
in the spin formalism, is the analog of spontaneous magnetization. 
The point $\alpha_c$ also marks the transition from persistence (for 
$\alpha>\alpha_c$) to anti-persistence ($\alpha<\alpha_c$) 
of the game's time series.


Let us start from a sharp definition of the model:
We use $+$ and $-$ to denote the two possible actions, 
so that a generic action is a sign.
At each time $t$, the information available to each agent
is the string $\mu_t=(\chi_{t-1},\ldots,\chi_{t-M})$ 
of the last $M$ actions taken by the minority. 
This, in our notation is a string of $M$ {\em minority signs} 
$\chi_{t-k}\in\{\pm 1\}$. There are $P=2^M$ possible 
such strings, which we shall label, by an index $\mu=1,
\ldots,P$\cite{notah}. The index $\mu_t$ 
corresponding to $(\chi_{t-1},\ldots,\chi_{t-M})$ shall
be called the present {\em history}, for short. For each history $\mu$, a strategy 
$a$ specifies a fixed action $a^\mu$ .
Each agent $i=1,\ldots,N$ has $S=2$ strategies, denoted by
$a_{\pm,i}$, which are randomly drawn from the 
set of all $2^P$ possible strategies (the  generalization 
to $S>2$ strategies will be discussed below). 
We define
\[
\omega_i^\mu=\frac{a_{+,i}^\mu+a_{-,i}^\mu}{2},~~~
\xi_i^\mu=\frac{a_{+,i}^\mu-a_{-,i}^\mu}{2}
\]
so that the strategies of agent $i$ can be written as
$a_{s_i,i}^\mu=\omega_i^\mu+s_i\xi_i^\mu$ with $s_i=\pm 1$. If $\omega_i^\mu\ne 0$, then $\xi_i^\mu=0$ (and viceversa) and the player always takes the decision $\omega_i^\mu$ whenever the history is $\mu$.
The current best strategy of agent $i$, which she shall adopt
at time $t$, is that which has the highest cumulated payoff.
Let us define $\Delta_{i,t}\equiv U^{(+)}_{i,t}-U^{(-)}_{i,t}$ 
as the difference between the cumulated payoffs $U^{(\pm)}_{i,t}$
of strategies $+$ and $-$ for agent $i$ at time $t$.
Therefore her choice is given by 
\be
s_i=\sign \Delta_{i,t}
\label{sit}
\ee
where ties ($\Delta_{i,t}=0$) are broken by coin tossing.
The difference in the population of agents choosing the $+$
and the $-$ sign, at time $t$, is then
\be
A_t=\sum_{i=1}^N a_{s_i,i}^{\mu_t}=
\Omega^{\mu_t}+\sum_{i=1}^N \xi_i^{\mu_t}s_i
\ee
where $\Omega^\mu=\sum_i\omega_i^\mu$.
The sign chosen by the minority gives the
{\em minority sign} at time $t$
\be
\chi_t = -\sign A_t
\label{supd}
\ee
and this determines the new history $\mu_{t+1}$ which
corresponds to the string $(\chi_t,\ldots,\chi_{t-M+1})$\cite{notah}. 
Finally, each agent $i$ rewards those of her strategies which have 
predicted the right sign ($a_{s,i}^{\mu_t}=\chi_t$) updating the
cumulated payoffs $U^{(\pm)}_{i,t+1}=U^{(\pm)}_{i,t}+a_{\pm,i}^{\mu_t}
\chi_t$. This implies that the cumulated payoff difference 
$\Delta_{i,t}$ is updated according to
\be
\Delta_{i,t+1}=\Delta_{i,t}+2\chi_t\xi_i^{\mu_t}.
\label{dit}
\ee

Eqs. (\ref{sit}--\ref{dit}) update the
state $\{\mu_{t},\Delta_{i,t}\}$ of the system from $t$
to $t+1$. With an initial condition (e.g. $\mu_0=1$,
$\Delta_{i,0}=0$, $\forall i$) the dynamics of the
MG is completely specified. The ``quenched'' 
variables $\{\Omega^\mu,\xi_i^\mu\}$ play here the same role
as disorder in statistical mechanics\cite{neuralnets}. 

\begin{figure}
\centerline{\psfig{file=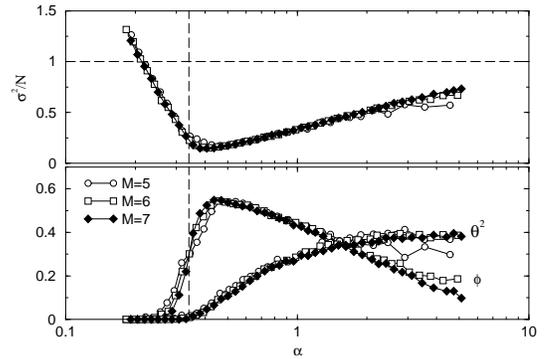,width=7cm}}
\caption{Top: $\sigma^2/N$ versus $\alpha=P/N$ for $P=2^M$ with 
$M=5,6$ and $7$. Bottom: $\theta^2$ and $\phi$ versus $\alpha$ for
the same system sizes $P$. The vertical dashed line is at 
$\alpha=0.34\approx\alpha_c$.}
\label{figphi}
\end{figure}

An important quantity in the MG is the 
variance $\sigma^2=\avg{A^2}$ of the difference
$A$ in the sizes of the two populations, where
$\avg{\cdot}$ is a time average in the stationary state 
of the process specified by Eqs. (\ref{sit}--\ref{dit}). 
The number of winners, at each time step, is 
$(N-|A|)/2\approx (N-\sigma)/2$ so 
that smaller fluctuations $\sigma^2$ correspond to 
larger global gain. A population of random agents
would yield $\sigma^2=N$. Numerical simulations 
\cite{CZ1,Savit,johnson} (see Fig. \ref{figphi}) show
that, for $\alpha=P/N$ large enough, agents with inductive reasoning manage to behave 
globally better (i.e. $\sigma^2<N$) than random 
agents whereas $\sigma^2>N$ 
for small $\alpha$ (see Fig. \ref{figphi}). 
However no singularity (and no order parameter) has been 
yet identified in order to locate a phase transition.

As shown in ref. \cite{cavagna}, to a good approximation one 
can neglect the coupling of the dynamics of $\Delta_{i,t}$ 
and $\mu_t$ and replace the dynamics of the latter by 
random sampling of the history space, 
i.e. ${\rm Prob}(\mu_t=\mu)=1/P$, $\forall \mu$. 
This simplifies considerably our discussion since then
\be
\sigma^2\simeq \frac{1}{P}\sum_{\mu=1}^P
\left(\Omega^\mu\right)^2+2\sum_{i=1}^N h_i\avg{s_i}+
\sum_{i,j=1}^NJ_{i,j}\avg{s_i s_j},
\label{s2}
\ee
where $\avg{\cdot}$ stands for a time average and
\be
h_i=\frac{1}{P}\sum_{\mu=1}^P\Omega^\mu\xi_i^\mu,~~~~~
J_{i,j}=\frac{1}{P}\sum_{\mu=1}^P\xi_i^\mu\xi_j^\mu.
\label{hJdef}
\ee
The field $h_i$ measures the difference of correlation
of the two strategies with $\Omega^\mu$ whereas the coupling
$J_{i,j}$ accounts for the interaction between agents as well
as for agents self-interaction ($J_{i,i}$).
The structure of the couplings \req{hJdef} is reminiscent of 
neural networks models\cite{neuralnets} where $\xi_i^\mu$ play 
the role of memory patterns. This similarity
confirms the conclusion of refs.\cite{ZENews,Savit,CZ2} that the 
relevant parameter is the ratio $\alpha=P/N$ between the number 
of patterns and the number of spins. 

The key element which is at the origin of the behavior of 
the model is the fact that for each history $\mu$, there are agents
which always take the same decision. This gives rise to 
the time independent contribution $\Omega^\mu$ in $A$ 
which produces a bias in the value of $\chi_t$
whenever $\mu_t=\mu$. 
A measure of this bias, is given by the parameter
\be
\theta=\sqrt{\frac{1}{P}\sum_{\mu=1}^P\avg{\chi|\mu}^2}
\label{theta}
\ee
where $\avg{\chi|\mu}$ is the conditional average of 
$\chi_t$ given that $\mu_t=\mu$. Loosely speaking, 
$\theta$ measures the presence of information or arbitrages 
in the signal $\chi_t$. If $\theta >0$ an agent with 
strategies of ``length'' $M=\log_2 P$, can detect and
exploit this information if one of her's strategies is 
more correlated with $\avg{\chi|\mu}$ than the other. 
More precisely, we observe 
that if $v_i\equiv\avg{\Delta_{i,t+1}-\Delta_{i,t}}\not = 0$ 
then $\Delta_{i,t}\simeq v_i t$ grows linearly with time, and
the agent's spin will always take the value $s_i=\sign v_i$. 
We shall call this a {\em frozen} agent, since her spin 
variable is frozen. We find 
\be
v_i=\avg{\chi_t\xi_i^{\mu_t}}\simeq \frac{1}{P}\!\sum_{\mu=1}^P
\avg{\chi |\mu}\xi_i^\mu
\propto -h_i -\sum_{j=1}^N J_{i,j} \avg{s_j}
\label{vi}
\ee
where the last equation relies on an expansion 
of $\avg{\chi|\mu}$ to linear order in 
$A$\cite{notalin}.

It is instructive to consider first the case where other agents choose by coin tossing (i.e. $\avg{s_j}=0$ 
for $j\ne i$) so that $v_i\propto - h_i - J_{i,i}\avg{s_i}$.
If $v_i\neq 0$ then $s_i=\sign v_i=-\sign(h_i + 
J_{i,i}\avg{s_i})$. But this last equation has a solution
only if $|h_i|>J_{i,i}$ whereas otherwise
$|\avg{s_i}|<1$ and $v_i=0$. Note that $J_{i,i}\simeq 1/2$
and that $h_i$ can be approximated by a gaussian variable 
with zero average and variance $(4\alpha)^{-1}$. This means
that $|h_i|\ll J_{i,i}$ for $\alpha\gg 1$, which implies that
most agents have $\avg{s_i}\approx 0$ in this limit and
we can indeed neglect agent--agent interaction. This allows to
compute the probability for an agent to be frozen 
\be
\phi=P\{|h_i|>J_{i,i}\}\propto e^{-\alpha/2},
\label{large}
\ee
for $\alpha\gg 1$. Numerical simulations show that 
$\phi\propto e^{-(0.37\pm 0.02)\alpha}$
indeed decays exponentially. As $\alpha\to\infty$, the random agents 
limit is attained because $\avg{s_i}\to 0$ for all $i$ and 
$\avg{s_is_j}=\avg{s_i}\avg{s_j}$ for $i\not = j$. By 
Eq. \req{s2} we find $\sigma^2=\sum_\mu(\Omega^\mu)^2/P+
\sum_i J_{i,i}\simeq N$.

The same argument applies in general, with the difference that 
the ``bare'' field $h_i$ must be replaced by the ``effective''
field $\tilde{h}_i=h_i+\sum_{j\ne i} J_{i,j}\avg{s_j}$. 
In order for agent $i$ to get frozen, 
her effective field $\tilde h_i$ must ovecome the self 
interaction $J_{i,i}$, i.e. 
$|\tilde h_i|> J_{i,i}\simeq 1/2$. If this condition is
met, $s_i=-\sign \tilde h_i$. It can also be shown
that a frozen agent will, on average, receive a larger
payoff than an unfrozen agent\cite{forthcoming}. 
Loosely speaking, one can say that a frozen agent has a 
{\em good} and a {\em bad} strategy and the good one remains better 
than the bad one even when she actually uses it.
On the contrary, unfrozen agents have two strategies each of which 
seems better than the other when it is not adopted. In this sense, 
symmetry breaking in $\avg{\chi|\mu}$ induced a sort of breakdown 
in the {\em a priori} equivalence of agents' strategies. 

A quantitative analysis of the fully interacting system 
shall be presented elsewhere\cite{forthcoming}. For the time 
being we shall discuss the behavior of the system on the 
basis of extensive numerical simulations.
Fig. \ref{figphi} reports the behavior of $\theta$, $\phi$
and $\sigma^2$ as functions of $\alpha$ for several values 
of $P$.
As $\alpha$ decreases, i.e. as more and more agents join 
the game, the arbitrages opportunities, as measured by 
$\theta$ decrease. In loose words, agents' exploitation of 
the signal $\Omega^\mu$ weakens its strength by screening
it with their adaptive behavior.
If the number $N$ of agents is small compared 
to the signal ``complexity'' $P=2^M$, agents exploit only 
partially the signal $\Omega^\mu$ whereas if $N\gg P$ then 
$\Omega^\mu$ is completely screened by agents' behavior and 
$\theta=0$. As Figure \ref{figphi} shows, the parameter $\theta$ 
displays the characteristic behavior of an order parameter 
with a singularity at $\alpha_c\approx 0.34$. Accordingly 
also the fraction $\phi$ of frozen agents drops to zero as
$\alpha\to\alpha_c^+$. 
The comparison between different system sizes in Fig. 
\ref{figphi} strongly suggests that $\phi$ drops 
discontinuously to zero at $\alpha_c$ (and it also gives
the value of $\alpha_c$). The vanishing of
$\phi$ is clearly a consequence of the fact that $\theta$ also 
vanishes at $\alpha_c$. 
Indeed if $\avg{\chi|\mu}=0$ for all $\mu$, by Eq. \req{vi}, 
also $v_i=0$ for all $i$, so that $\Delta_{i,t}$ remains 
bounded and $|\avg{s_i}|<1$.

The transition can also be understood in terms of the 
variables $\Delta_{i,t}$ as an ``unbinding'' transition  
as $\alpha\to\alpha_c^-$:
For $\alpha<\alpha_c$ a ``bound state'' exists with
finite $\Delta_{i,t}$, which corresponds to the fact that the 
equations $v_i=0$, $i=1,\ldots,N$ admit a solution with 
$|\avg{s_i}|<1$, $\forall i$\cite{forthcoming} (only 
$P$ of the equations $v_i=0$ are linearly independent). 
For $\alpha>\alpha_c$ this is no longer true and 
the population separates: a fraction $\phi$ of 
variables $\Delta_{i,t}$ acquire a constant ``velocity'' 
$v_i\not = 0$ (with $|\avg{s_i}|=1$) whereas for the
remaining agents $v_i=0$, $\Delta_{i,t}$ remains
bounded and $|\avg{s_i}|<1$. 

It is suggestive to observe that
$v_i\propto-\frac{\partial {\sigma^2}}{\partial s_i}$ so 
that the dynamics of the minority game is actually similar to 
a spin dynamics with hamiltonian $\sigma^2$. Indeed either
the spin is frozen in the direction which minimizes 
$-s_i v_i(s_i)$, or its average $\avg{s_i}$
is such that $v_i=0$. This then explains why cooperation 
occurs in the MG. A closer analysis, to be reported 
elsewhere\cite{forthcoming}, reveals that indeed the stationary
state of the MG is described by the ground state properties 
of an Hamiltonian very similar to $\sigma^2$.
Finite size scaling suggests that $\sigma^2$ has a 
minimum at $\alpha_c$ with a discontinuity in its derivative
(see Fig. \ref{figphi}). These conclusions are indeed confirmed 
by exact results\cite{forthcoming}.
It is worth to stress, however, that the qualitative aspects of 
the transition are already captured at the simple level of 
approximation of Eq. \req{vi}. 

\begin{figure}
\centerline{\psfig{file=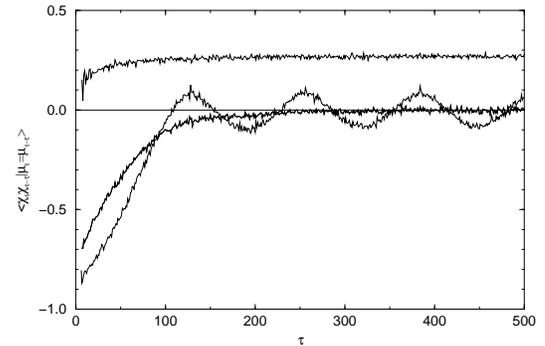,width=7cm}}
\caption{Temporal correlation of $\chi_t$ on the same history,
$\avg{\chi_t\chi_{t+\tau}| \mu_t=\mu_{t+\tau}}$, averaged over
all histories versus $\tau$ ($10^6$ iterations, $M=6$, 
$\alpha=0.5$, $0.22$, $0.1$)}
\label{figsigmat}
\end{figure}

Let us go back to Fig. \ref{figphi}.
Above $\alpha_c$ agents do not fully exploit the information 
$\Omega^\mu$ and, as a result, $\avg{\chi|\mu}\not =0$. 
Figure \ref{figsigmat} shows that 
$\chi_t$ shows persistence in time, in the sense that when 
$\mu_t=\mu_{t+\tau}$ the minority signs $\chi_t$ and
$\chi_{t+\tau}$ tend to be the same. This persistence 
disappears, $\avg{\chi_t\chi_{t+\tau}|\mu_t=\mu_{t+\tau}}\to 0$ 
as $\alpha$ decreases and it turns into anti-persistence for 
smaller $\alpha$. The oscillatory behavior in Fig. \ref{figsigmat} 
has indeed period $2P$ which means that typically when the 
population comes again on the same history $\mu$ it tends 
to do the opposite of what it did the time before. 
Even if finite size effects do not allow a definite conclusion,
it is quite likely that this change in time correlations also 
occurs at $\alpha_c$\cite{forthcoming}. 
Time correlations, even though of opposite
nature, are present both above and below $\alpha_c$. 
These are like arbitrages in a market 
which could be exploited by agents. In this
sense the market is {\em efficient}, i.e. arbitrage free,
only for $\alpha=\alpha_c$.

The same qualitative behavior is expected when agents have $S>2$ 
strategies. Again for a given history $\mu$ it may happen 
that all the $S$ agent's $i$ strategies prescribe the same 
action: agent $i$ will do that action no matter what strategy 
she has choosen. As $S$ increases, this will occur for a 
smaller and smaller number of histories (more precisely
with a probability $2^{1-S}$). This shall correspond to 
a weaker signal $\Omega^\mu$ which is in complete agreement 
with the observation \cite{Savit,CZ2} of shallower 
features for larger $S$. Note that, for each agent it
would be rewarding to increase the number of strategies
because they would have more chances to outguess $\chi_t$.
At the same time, if all agents increase $S$ the game 
becomes less rewarding for all of them, at least for
$\alpha>\alpha_c$. This situation is typical of games, such
as the {\em tragedy of commons}, where many agents interact
through a global resource\cite{hardin}.

The condition $v_i=0$ for the bound state in the symmetric phase
involves $P$ equations with $(S-1)N$ variables. This suggests
that in general the scaling parameter is $\alpha=P/[(S-1)N]$. 
The curve $\sigma^2/N$ as a function of $\alpha=P/[(S-1)N]$ 
collapse remarkably well one on the other for $\alpha\le\alpha_c$ 
(especially for $S>2$) but not for $\alpha>\alpha_c$ (e.g. in
the large $\alpha$ behavior $\phi\propto e^{-C(S)\alpha}$ we
found $C(2)\approx 0.37$, $C(3)\approx 1.50$ 
and $C(4)\approx 2.90$).

Our approach also implies that no coordination is
possible if agents have
$S=2$ opposite strategies ($a^\mu_{+,i}=-a^\mu_{-,i}$)
because then $\Omega^\mu =  0$.
Numerical simulations 
show that indeed $\sigma^2 \ge N$ for all $\alpha >0$ in this case.

The same qualitative behavior also occurs in a wide 
range of related models. First, total freezing
occurs in majority models. Note indeed that changing the sign of 
Eq. (\ref{supd}) would also change the sign in Eq. \req{vi}.
In particular the self-interaction $J_{i,i}$ changes sign so 
that it becomes favorable for each agent to stick to 
only one strategy anyway. The model is therefore trivial.
More interesting models are obtained keeping the ``frustration'' 
effects of the MG but changing the definition
of payoffs in Eq. \req{dit}. It can be shown\cite{forthcoming}
that the phase transition and the large $\alpha$ behavior
are quite robust features of minority games (see e.g. 
\cite{notalin}).

In summary we find that a phase transition occurs in the
minority game. The cooperative phase ($\alpha>
\alpha_c$) is characterized by the presence of a fraction 
$\phi$ of frozen agents (who use only one strategy),
unexploited arbitrages ($\avg{\chi|\mu}\not = 0$) 
and persistence in the global signal $\chi_t$.
In the symmetric phase ($\alpha<\alpha_c$) inductive
dynamics is inefficient: agents adopt strategies when 
they are no more good. There is no arbitrage (for 
strategies of length $M$) to exploit and the signal 
shows anti-persistence.



We acknowledge Y.-C. Zhang for enlightening discussions, 
useful suggestions and for introducing us to the Minority 
Game. This work was partially supported by
Swiss National Science Foundation Grant Nr 20-46918.98.

\begin{thebibliography}{99}
\bibitem{notah} An example of labeling is binary
coding. $\mu_{t+1}$ is then given by:
$\mu_{t+1}=[2\mu_t+\frac{1+\chi_t}{2}]\;{\rm mod}\; P$.
\bibitem{cavagna} A. Cavagna, Phys. Rev. {\bf E 59}, R3787 (1999)
has shown that replacing the dynamics of $\mu_t$ by random 
sampling does not change, to a very good approximation, the
behavior of the model. The assumption that each history $\mu$ 
is sampled with the same frequency is strictly true only in
the symmetric phase $\alpha<\alpha_c$ but it holds qualitatively
also for $\alpha>\alpha_c$.
\bibitem{notalin} This approximation is justified by the fact 
that the ``linear'' minority model obtained replacing Eq. 
\req{dit} by $\Delta_{i,t+1}=\Delta_{i,t}-A_t 
\xi_i^{\mu_t}$ gives results very similar to those presented 
here. An analytic derivation is also possible from the
identity $\avg{\chi|\mu}=2{\rm Prob}(A^\mu<0 
|\mu)-1$\cite{forthcoming}.
\end{thebibliography}
\end{document}